\documentclass[aps,prd,amsmath,twocolumn,amssymbaps,showpacs]{revtex4-2}
\usepackage{graphicx}  
\usepackage{dcolumn}   
\usepackage{bm}        
\usepackage{amssymb}   
\usepackage{amsmath}   
\usepackage{bbm}
\usepackage{draftcopy}
\usepackage{relsize} 
\usepackage{xfrac}    
\usepackage{slashed}
\usepackage{color}
\usepackage{footnote}
\definecolor{dgreen}{cmyk}{1.,0.,1.,0.2}        
\definecolor{orange}{cmyk}{0.,0.353,1.,0.}    

\usepackage[bookmarks]{hyperref}

\newcommand{\di}{{\rm d}}

\newcommand{\be}{\begin{equation}}
\newcommand{\ee}{\end{equation}}                                                                               
\newcommand{\bea}{\begin{eqnarray}}
\newcommand{\eea}{\end{eqnarray}}

\begin{document}
\title{Single-flavor heavy baryons in a strong magnetic field}

\author{Gaoqing Cao$^{1,2}$ and Shuang Wu$^1$}
\affiliation{1 School of Physics and Astronomy, Sun Yat-sen University, Zhuhai 519082, China\\
2 Guangdong Provincial Key Laboratory of Quantum Metrology and Sensing, Sun Yat-Sen University, Zhuhai 519082 
China}
\date{\today}

\begin{abstract}
In this work, we study the properties of single-flavor heavy baryons, $\Omega_{\rm ccc}$ and $\Omega_{\rm bbb}$, in a strong magnetic field. For that sake, we simply treat the baryons as quark-diquark two-body systems, and a systematic formalism is developed to deal with two-body Schr$\ddot{\text o}$dinger equations in a magnetic field. It is found that: 1. The orbital properties of $\Omega_{\rm bbb}$ are almost not affected by the magnetic field. 2. $\Omega_{\rm ccc}$ is more tightly bound in the presence of a magnetic field. 3. The magnetic-spin effect dominates over the magnetic-orbital effect. Applying to peripheral heavy ion collisions, $\Omega_{\rm ccc}$ is much better than $\Omega_{\rm bbb}$ to explore the magnetic effect, and the discovery of $\Omega_{\rm ccc}$ could be more promising.
\end{abstract}


\maketitle

\section{Introduction}
In relativistic heavy ion collisions (HICs), a hot and dense quark-gluon plasma (QGP) can be produced and the properties have been well explored by employing different kinds of probes: photon, dilepton, jet, heavy hadrons, etc.~\cite{Yagi:2005yb}. Specifically, jet and heavy hadrons are usually produced at early stage of the collisions and evolve through the whole stages of QGP, thus their detective spectra could reflect the variations of the QGP environment. On the other hand, compared to those in p-p collision, the production rates of heavy hadrons could be greatly amplified in relativistic HICs, so HICs are considered ideal experiments to look for new heavy hadrons, such as $\Omega_{\rm ccc}$ and $\Omega_{\rm bbb}$~\cite{He:2014tga,Zhao:2023qww}.

 Recent years, new circumstances emerge in peripheral HICs. For example, the magnetic field can be as large as $1~{\rm GeV}^2$ at LHC energy~\cite{Voronyuk:2011jd,Bzdak:2011yy,Deng:2012pc} and the angular velocity is $10^{22}~s^{-1}$ at RHIC energy~\cite{STAR:2017ckg}, though they both decrease with time~\cite{Zhao:2020jqu}. The properties of heavy mesons have been extensively explored in the presence of strong electromagnetic (EM) field and fast rotation, see the review Ref.~\cite{Zhao:2020jqu}. In the presence of a magnetic field, it was found that mixing would be introduced to the scalar and vector quarkonia, and the mass of scalar-dominate meson decreases with magnetic field while that of vector-dominate meson increases with it~\cite{Guo:2015nsa}. In the presence of an electric field, all the quarkonium masses would mainly tend to decrease due to the deconfinement effect introduced~\cite{Xiang:2024wxh}. In the presence of rotation, all the quarkonia are expected to be more easily dissociated due to the effective chemical potential introduced~\cite{Jiang:2016wvv,Chen:2015hfc,Cao:2021rwx,Cao:2021jjy}. Moreover, the interplay among EM fields and rotation is very interesting~\cite{Chen:2020xsr}: For a strong electric field and a finite rotation, $J/\psi_0$ would transit from strong bound state to EM and rotational bound state with increasing magnetic field.
 
The properties of baryons have not been well explored in the presence of EM field or rotation, because the systems are basically anisotropy and the well-developed hyper geometric method~\cite{Zhao:2020jqu} might not easily apply. As a starting point, we first consider the simplest single-flavor heavy baryons, which are more symmetric, in the presence of a magnetic field. And this work is organized as follows: In Sec.~\ref{MF}, a whole formalism is developed to deal with few-body Schr$\ddot{\text o}$dinger equations in a magnetic field, where Sec.~\ref{TBSE} is devoted to representing the basic two-body Schr$\ddot{\text o}$dinger equations in center of mass and relative coordinates and Sec.~\ref{baryons} to discussing baryons based on a self-consistent quark-diquark approximation. In Sec.~\ref{num}, the Schr$\ddot{\text o}$dinger equations are solved numerically for two single-flavor heavy baryons, $\Omega_{\rm ccc}$ and $\Omega_{\rm bbb}$. Finally, a summary is given in Sec.~\ref{summary}.

\section{Schr$\ddot{\text O}$dinger equations in a constant magnetic field}\label{MF}
In this section, we try to establish a formalism to study the effect of a constant magnetic field on heavy hadrons, composed of charm or bottom (anti-)quarks, by solving the few-body Schr$\ddot{\text o}$dinger equations. If only two-body forces are considered, the $n$-body hadronic wave function $\Psi({\bf r}_1,\dots,{\bf r}_{\rm n})$ satisfies the following static Schr$\ddot{\text o}$dinger equation:
\bea
\left[\sum_{\rm j=1}^n{(\hat{\bf p}^{\rm j}-q^{\rm j} {\bf A}({\bf r}_{\rm j}))^2\over2m_{\rm j}}+U({\bf r}_1,\dots,{\bf r}_{\rm n})-E\right]\Psi=0,\label{SE1}
\eea
where the kinetic momentum of (anti-)quark $j$ is defined as
 \bea
 \hat{p}_{\rm i}^{\rm j}-q^{\rm j} A_{\rm i}({\bf r}_{\rm j})\equiv -i{\partial\over\partial r_{\rm j}^{\rm i}}-q^{\rm j}A_{\rm i}({\bf r}_{\rm j}).
 \eea
 and the total potential is given by the sum of all two-body potentials, $V({\bf r}_{\rm jj'})$ with  ${\bf r}_{\rm jj'}\equiv{\bf r}_{\rm j}-{\bf r}_{\rm j'} $ the relative displacement, as
\bea
U({\bf r}_1,\dots,{\bf r}_{\rm n})\equiv{1\over2}\sum_{j,j'=1,\dots,n}^{j\neq j'}V({\bf r}_{\rm jj'}),
\eea
Note that magnetic-spin coupling might also be important for the study of hadronic properties~\cite{Zhao:2020jqu}, but we will temporarily suppress that in this section, since it is not relevant to the derivations in coordinate space given below. Without loss of generality, we choose the magnetic field to be along $z$-direction and work in the symmetric gauge, that is, ${\bf A}({\bf r}_{\rm j})=B(-y_{\rm j}/2,x_{\rm j}/2,0)$. The two-body potential between quark and antiquark,  $V_{\rm q \bar{q}}({\bf r})$, has been well explored for zero and finite temperatures and usually takes the isotropic form~\cite{Lafferty:2019jpr}
 \bea
V_{\rm q \bar{q}}(r)\!=\!-\alpha_{\rm s}\!\left(m_{\rm D}\!+\!{e^{-m_{\rm D}r}\over r}\right)\!+\!{\sigma\over m_{\rm D}}\!\!\left[2\!-\!(2\!+\!m_{\rm D}r){e^{-m_{\rm D}r}}\right]\label{Vqbq}\nonumber\\
\eea
with the Coulomb coupling constant $\alpha_{\rm s}=0.4105$ and the string tension $\sigma=0.2~ {\rm GeV}^2$~\cite{Hu:2022ofv}. More refined, there should be spin-spin interaction terms in the potential~\cite{Zhao:2020jqu}, but we will neglect them for simplicity in the following. According to both perturbative QCD calculations and quark model, the two-body potential between quarks is just one half of that, that is,  $V_{\rm q q}({\bf r})=V_{\rm q \bar{q}}(r)/2$~\cite{Rothkopf:2019ipj}.

Furthermore, for reader's reference, both the effects of finite temperature and baryon density can be introduced by simply modifying the Debye mass and accordingly shifting the energy scale from $2\pi T$ to $\Lambda\equiv 2\sqrt{(\pi T)^2+\mu_{\rm B}^2}$, we have~\cite{Lafferty:2019jpr,Huang:2021ysc,Schneider:2003uz}
\bea
&&m_{\rm D}=\sqrt{m_{\rm D}^2(T)+m_{\rm D}^2(\mu_{\rm B})}, \nonumber\\
&&m_{\rm D}(T)=g_\Lambda T\sqrt{{N_{\rm c}\over3}\!+\!{N_{\rm f}\over6}}+{N_{\rm c}Tg_\Lambda^2\over 4\pi}\log\left({1\over g_\Lambda}\sqrt{{N_{\rm c}\over3}\!+\!{N_{\rm f}\over6}}\right)\nonumber\\
&&\qquad\qquad\ +\kappa_1 T g_\Lambda^2 +\kappa_2 T g_\Lambda^3+\kappa_3T g_\Lambda^5,\nonumber\\
&&m_{\rm D}^2(\mu_{\rm B})=g_\Lambda^2N_{\rm f}{\mu_{\rm B}^2\over18\pi^2}
\eea
with $\kappa_1 =0.6, \ \kappa_2=-0.23$ and $\kappa_3=-0.007$.
Here, the running coupling $g_\Lambda$ is determined according to the differential equation for $\alpha\equiv g^2/4\pi^2$~\cite{Vermaseren:1997fq},
\bea
&&{d \alpha\over d \ln \mu^2}=-(\beta_0 \alpha^2+\beta_1 \alpha^3+\beta_2\alpha^4+\beta_3\alpha^5),\\
&&\beta_0=2.750-0.1667N_{\rm f}, \ \ \beta_1=6.375-0.7917N_{\rm f}, \nonumber\\
&&\beta_2=22.32-4.369N_{\rm f}+0.09404N_{\rm f}^2,\nonumber\\
&&\beta_3=114.2-27.13N_{\rm f}+1.582N_{\rm f}^2+0.005857N_{\rm f}^3
\eea
with the initial condition $\alpha(\Lambda_{\rm QCD})=\alpha_{\rm s}$. For a realistic study of heavy hadrons in heavy ion collisions, we take $N_{\rm c}=N_{\rm f}=3$ and $\Lambda_{\rm QCD}=0.2~ {\rm GeV}$, and assume the heavy flavors to be not thermalized themselves. The latter explains why there is no explicit temperature or baryon density effect in the Schr$\ddot{\text o}$dinger equation \eqref{SE1}. So, the medium effect is introduced into the Schr$\ddot{\text o}$dinger equation solely through the Debye mass $m_{\rm D}$, and we will take different $m_{\rm D}$ for comparison in the end.

\subsection{Two-body Schr$\ddot{\text o}$dinger equation}\label{TBSE}
 Two-body Schr$\ddot{\text o}$dinger equation is the basic one for solving the $n$-body Schr$\ddot{\text o}$dinger equation, since the latter can be approximately separated into several two-body problems. For example, baryons, composed of three quarks, can be treated as a quark-diquark two-body system with diquark itself a quark-quark two-body system. So, we will focus on the two-body Schr$\ddot{\text o}$dinger equation first in the following. Set $n=2$, \eqref{SE1} is reduced to
 \bea
\left[{(\hat{p}_{\rm i}^1\!-\!q^1A_{\rm i}({\bf r}_1))^2\over2m_1}+{(\hat{p}_{\rm i}^2\!-\!q^2A_{\rm i}({\bf r}_2))^2\over2m_2}\!+\!V({\bf r}_{21})\!-\!E\right]\!\Psi=0,\nonumber\\\label{SE2}
\eea
where the Einstein summation conventions should be understood for the coordinate indices ${\rm i}=x,y,z$.

Inspired by the Schwinger phase in a magnetic field, if we redefine the wave function as
\bea
\Psi({\bf r}_1,{\bf r}_2)=e^{i\,{q^{21}\over2}\Phi({\bf r}_1,{\bf r}_2)}\tilde{\Psi}({\bf r}_1,{\bf r}_2),
\eea
 with $q^{21}\equiv q^2-q^1$ the charge difference and
 \bea
 \Phi({\bf r}_1,{\bf r}_2)\equiv \int_{{\bf r}_1}^{{\bf r}_2} \left[A_\mu({\bf r})+{1\over2}F_{\mu\nu}(r-{r}_1)^\nu\right]\di r^\mu
 \eea
the charge-free Schwinger phase,  the Schr$\ddot{\text o}$dinger equation \eqref{SE2} can be reduced to a more symmetric form
\bea
&&\!\!\!\!\left[{(\hat{p}_{\rm i}^1\!-\!{Q\over2}A_{\rm i}({\bf r}_1)\!+\!{q^{21}\over4}F_{\rm i\nu}{r}_{21}^\nu)^2\over2m_1}+\right.\nonumber\\
&&\left.{(\hat{p}_{\rm i}^2\!-\!{Q\over2}A_{\rm i}({\bf r}_2)\!+\!{q^{21}\over4}F_{\rm i\nu}{r}_{21}^\nu)^2\over2m_2}\!+\!V({\bf r}_{21})\!-E\right]\tilde{\Psi}=0
\eea
with the total charge $Q=q_1+q_2$. Since the potential only depends on the relative coordinate, ${\bf r}_{21}$, it is more convenient to rewrite the equation in the form of two kinds of motions: center of mass motion with total mass $M\equiv\sum_{\rm j} m_{\rm j}$ and coordinates ${\bf R}=\sum_{\rm j} \tilde{m}_{\rm j} {\bf r}_{\rm j}\ (\tilde{m}_{\rm j}\equiv{m_{\rm j}\over M})$, and relative motion with coordinates ${\bf r}_{\rm jj'}$.  For a two-body system, 
\bea
&&\tilde{m}_1+\tilde{m}_2=1,\\
&& {\bf R}={\tilde{m}_1{\bf r}_1+\tilde{m}_2{\bf r}_2},\\
&& {\bf r}_1={\bf R}-\tilde{m}_2{{\bf r}_{21}},\\
&& {\bf r}_2={\bf R}+\tilde{m}_1{{\bf r}_{21}},
\eea
 then the Schr$\ddot{\text o}$dinger equation becomes
 \begin{widetext}
\bea
\left[{({\tilde{m}_1\hat{p}_{\rm i}^{R}+\bar{A}_{\rm i}}-\hat{p}_{\rm i}^{r}+{Q\over4}A_{\rm i}({\bf r}_{21}))^2\over 2\tilde{m}_1M}+{({\tilde{m}_2\hat{p}_{\rm i}^{R}+\bar{A}_{\rm i}}+\hat{p}_{\rm i}^{r}-{Q\over4}A_{\rm i}({\bf r}_{21}))^2\over 2\tilde{m}_2M}+V({\bf r}_{21})-E\right]\tilde{\Psi}({\bf R},{\bf r}_{21})&=&0,\nonumber\\
\left[{(\hat{p}_{\rm i}^{R}+2\bar{A}_{\rm i})^2\over 2M}+{(\hat{p}_{\rm i}^{r}-{Q\over4}A_{\rm i}({\bf r}_{21})+({\tilde{m}_1-\tilde{m}_2})\bar{A}_{\rm i})^2\over 2\mu}+V({\bf r}_{21})-E\right]\tilde{\Psi}({\bf R},{\bf r}_{21})&=&0
\eea
with $\hat{p}_{\rm i}^{r}\equiv -i{\partial\over\partial r_{21}^{\rm i}}$, $\bar{A}_{\rm i}\equiv -{Q\over2}A_{\rm i}({\bf R})+{q^{21}\over4}F_{\rm i\nu}{r}_{21}^\nu-{\tilde{m}_1-\tilde{m}_2\over2}{Q\over2}A_{\rm i}({\bf r}_{21})$, and the relative reduced mass ${\mu}=m_1m_2/(m_1+m_2)$. Note that $\hat{p}_{\rm i}^{R}, \bar{A}_{\rm i}, \hat{p}_{\rm i}^{r}$ and $A_{\rm i}({\bf r}_{21})$ are commutative with each other. 

Since $A_{\rm i}({\bf r}_{21})=-{1\over2}F_{\rm i\nu}{r}_{21}^\nu=-{1\over2}({\bf B}\times {\bf r}_{21})_{\rm i}=A_{\rm i}({\bf r}_{2})-A_{\rm i}({\bf r}_{1})$ in the symmetric gauge, the Schr$\ddot{\text o}$dinger equation is explicitly
\bea
\left[{(\hat{p}_{\rm i}^{R}-{Q}A_{\rm i}({\bf R})+{\bar{q}'\over2}F_{\rm i\nu}{r}_{21}^\nu)^2\over 2M}+{(\hat{p}_{\rm i}^{r}+{q''\over2}F_{\rm i\nu}{r}_{21}^\nu-({\tilde{m}_1-\tilde{m}_2}){Q\over2}A_{\rm i}({\bf R}))^2\over 2\mu}+V({\bf r}_{21})-E\right]\tilde{\Psi}({\bf R},{\bf r}_{21})=0\label{SE3}
\eea
with the effective charges $\bar{q}'=q^{21}+{\tilde{m}_1-\tilde{m}_2\over2}Q$ and $q''=\tilde{m}_1^2q^2+\tilde{m}_2^2q^1$. As $A_{\rm i}({{\bf R}})=-{1\over2}F_{i\nu}{R}^\nu$, if we redefine $\tilde{\Psi}({\bf R},{\bf r}_{21})=e^{-i({\tilde{m}_1-\tilde{m}_2}){{Q}\over4}F_{\mu\nu}{R}^\nu r_{21}^\mu}\bar{\Psi}({\bf R},{\bf r}_{21})$, the Schr$\ddot{\text o}$dinger equation \eqref{SE3} can be further simplified to 
\bea
\left[{(\hat{p}_{\rm i}^{R}-{Q}A_{\rm i}({\bf R})+{q'}F_{\rm i\nu}{r}_{21}^\nu)^2\over 2M}+{(\hat{p}_{\rm i}^{r}+{q''\over2}F_{\rm i\nu}{r}_{21}^\nu)^2\over 2\mu}+V({\bf r}_{21})-E\right]\bar{\Psi}({\bf R},{\bf r}_{21})=0\label{Rr}
\eea
with ${q'}\equiv {\tilde{m}_1q^2-\tilde{m}_2q^1}$; or alternatively if we redefine $\tilde{\Psi}({\bf R},{\bf r}_{21})=e^{i{\bar{q}'\over4}F_{\mu\nu}{R}^\nu r_{21}^\mu}\bar{\Psi}({\bf R},{\bf r}_{21})$, the Schr$\ddot{\text o}$dinger equation \eqref{SE3}  can be reduce to
\bea
\left[{(\hat{p}_{\rm i}^{R}-{Q}A_{\rm i}({\bf R}))^2\over 2M}+{(\hat{p}_{\rm i}^{r}+{q''\over2}F_{\rm i\nu}{r}_{21}^\nu-{q}'A_{\rm i}({\bf R}))^2\over 2\mu}+V({\bf r}_{21})-E\right]\bar{\Psi}({\bf R},{\bf r}_{21})=0.\label{Rr1}
\eea
\end{widetext}
So, we recognize that $q''$ is the reduced charge for the relative motion, which is actually evaluated in the same way as the reduced mass with the observation $\mu=\tilde{m}_1^2m_2+\tilde{m}_2^2m_1$.

In the case $q'\neq 0$, there is mixing between center of mass and relative motions and the Landau eigenstates are no longer the orthogonal basis for either motion. But we can still work on the complete Landau basis and expand over the corresponding eigenfunctions for the center of mass motion, 
\bea
\chi_{l,n}(\theta,\tilde{R}_\bot)\equiv\left[{|QB|\over2\pi}{n!\over(n+l)!}\right]^{1/2}e^{i l\theta}\tilde{R}_\bot^l e^{-\tilde{R}_\bot^2/2}L_{\rm n}^l(\tilde{R}_\bot^2)\nonumber\\
\eea
 with $\tilde{R}_\bot^2\equiv |QB|{R}_\bot^2/2$ the reduced transverse radius and $L_{\rm n}^l(\tilde{R}_\bot^2)\ (n\in \mathbb{N}, l\geq-n)$ the Laguerre polynomial. Of course, the mixing term in \eqref{Rr1} would give rise to nonorthogonal terms like $\chi_{l,n\pm1}(\theta,\tilde{R}_\bot)$, then the true eigenstates should be solved by diagonalizing the Hamiltonian matrix in Landau space. 
 
 There are several exceptions: For a single heavy flavor $n$-body system, ${\tilde{m}_1\over q^1}={ \tilde{m}_2\over q^2}$ between any two (quasi-)particles, so the center of mass motion and relative motion become effectively independent. If $M\gg \mu$, the first term in \eqref{Rr} can be simply neglected and the relative Schr$\ddot{\text o}$dinger equation is just
\bea
\left[{(\hat{p}_{\rm i}^{r}+{q''\over2}F_{\rm i\nu}{r}_{21}^\nu)^2\over 2\mu}+V({\bf r}_{21})-E\right]\bar{\Psi}({\bf r}_{21})=0.
\eea
The latter can be adopted as a rough approximation to study the charm-bottom two-body system in a not too strong magnetic field since $m_{\rm b}=3.64~m_{\rm c}$ and $\mu/M=0.17$. Note that in the limit $B\rightarrow 0$, this exactly gives the relative Schr$\ddot{\text o}$dinger equation for exploring the relative energy. 

\subsubsection{Equal-mass systems}\label{Em}
A single flavor system is an equal-mass system and a single flavor and anti-flavor system is also an equal-mass system. There is also charge symmetry between a flavor and a (anti-)flavor, that is, $q_1=\pm q_2$, so we will show how the Schr$\ddot{\text o}$dinger equation can be simplified and solved for such systems. Take $m_1=m_2\equiv m$, the Schr$\ddot{\text o}$dinger equation \eqref{Rr} is then reduced to
\bea
&&\left[{(\hat{p}_{\rm i}^{R}\!-\!{Q}A_{\rm i}({\bf R})\!+\!{q^{21}\over2}F_{\rm i\nu}{r}_{21}^\nu)^2\over 2M}+\right.\nonumber\\
&&\left.\qquad\qquad {(\hat{p}_{\rm i}^{r}\!+\!{Q\over8}F_{\rm i\nu}{r}_{21}^\nu)^2\over 2\mu}\!+\!V({\bf r}_{21})\!-\!E\right]{\Psi}=0.
\eea

For a flavor-anti-flavor system with charges $q_2=-q_1=q$, $Q=0$ and $q^{21}=2q$, so we have 
\bea
\left[{(\hat{p}_{\rm i}^{R}+{q}F_{\rm i\nu}{r}_{21}^\nu)^2\over 2M}+{(\hat{p}_{\rm i}^{r})^2\over 2\mu}+V({\bf r}_{21})-E\right]{\Psi}=0.
\eea
As the pseudomomentum $\hat{p}_{\rm i}^{R}$ commutates with the effective Hamiltonian, if we redefine ${\Psi}({\bf R},{\bf r}_{21})=e^{i {\bf P\cdot R}}\psi({\bf r}_{21})$, the relative Schr$\ddot{\text o}$dinger equation follows as
\bea
\!\!\!\!\!\!\!\!\left[{(P_{\rm i}\!+\!{q}F_{\rm i\nu}{r}_{21}^\nu)^2\over 2M}-{P_{\rm i}^2\over 2M}\!+\!{(\hat{p}_{\rm i}^{r})^2\over 2\mu}\!+\!V({\bf r}_{21})\!-E_{\rm r}\right]\psi=0.
\eea
This equation is exactly the same as that given in Ref.~\cite{Guo:2015nsa}. In heavy ion collisions, the longitudinal expansion is less significant, so we can take ${\bf P}$ to be along $x$-direction in order to maximize the effect. This provides a useful basis for further study of equal-mass tetraquark and pentaquark states in a constant magnetic field. 

For a single flavor two-body system, $Q=2q$ and $q^{21}=0$, so we have 
\bea
\left[{(\hat{p}_{\rm i}^{R}\!-\!{Q}A_{\rm i}({\bf R}))^2\over 2M}\!+\!{(\hat{p}_{\rm i}^{r}\!+\!{Q\over8}F_{\rm i\nu}{r}_{21}^\nu)^2\over 2\mu}\!+\!V({\bf r}_{21})\!-E\right]{\Psi}=0,\nonumber\\
\eea
where the relative charge can be found to be $Q/4=q/2$. As mentioned before, the center of mass motion and relative motion are independent in such a case, then we can redefine ${\Psi}({\bf R},{\bf r}_{21})=\psi({\bf r}_{21})\Phi({\bf R})$ to separate these two parts as
\bea
\left[{({\hat{p}_{\rm i}^R-{Q}A_{\rm i}({\bf R})})^2\over 2M}-(E-E_{\rm r})\right]\Phi({\bf R})&=&0,\label{SER}\\
\left[{(\hat{p}_{\rm i}^{r}+{Q\over8}F_{\rm i\nu}{r}_{21}^\nu)^2\over 2\mu}+V({\bf r}_{21})-E_{\rm r}\right]\psi({\bf r}_{21})&=&0.\label{SEr}
\eea
Here, $\Phi({\bf R})$ is the diquark global wave function. 

\subsection{Baryons as a quark-diquark system}\label{baryons}
As mentioned in the Introduction, the properties of heavy-flavor mesons have been well studied in the presence of a magnetic field. It was found that the total momentum is not a good quantum number even for a chargeless meson, and there are mixings among different spin states~\cite{Guo:2015nsa}. Here, we will just focus on single-flavor baryons, the properties of which has not been so widely explored in the presence of a magnetic field.  As described before, a baryon can be simply considered as a quark-diquark two-body system in a magnetic field, that is, quarks $1$ and $2$ form a independent diquark (see \eqref{SER}) which then couples with quark $3$. In such an approximation, we can apply the formalism developed in Sec.\ref{TBSE} twice to solve the relative energy of the baryon in principle. 

However, an additional approximation is needed since the third quark would introduce two more potentials that depend on both relative coordinates inside diquark and quark-diquark systems. Due to the symmetry between quarks $1$ and $2$, it is reasonable to assume the three quarks to mainly form an isosceles triangle with the apex angle $2\theta$ at the point of quark $3$, then $|{\bf r}_{31}|=|{\bf r}_{32}|=|\boldsymbol{\rho}|/\cos\theta$ with $\boldsymbol{\rho}\equiv{\bf r}_3-{\bf R}$ and $\boldsymbol{\rho}\cdot {\bf r}_{21}=0$.  For future use, the extra potentials related to quark $3$, $V({\bf r}_{31})$ and $V({\bf r}_{32})$, can be alternatively presented as the potentials depending on ${\bf r}_{21}$ and $\boldsymbol{\rho}$ as
\begin{widetext}
\bea
\!\!\!\!V_{12}({\bf r}_{21},\theta)&=&-\int {\bf f}_{31} \cdot d{\bf r}_{1}-\int {\bf f}_{32}\cdot d{\bf r}_{2}=\int V'\left(r_{32}\right)\sin\theta dr_{21}=\int V'\left({r_{21}\over2\sin\theta}\right)\sin\theta dr_{21}=2\sin^2\theta\ V\left({r_{21}\over2\sin\theta}\right),\\
V_{R3}(\boldsymbol{\rho},\theta)&=&-\int {\bf f}_\rho\cdot d\boldsymbol{\rho}=\int [V'(r_{31})+V'(r_{32})]\cos\theta d\rho=2\int V'\left({\rho\over\cos\theta}\right) \cos\theta d\rho=2\cos^2\theta\ V\left({\rho\over\cos\theta}\right)
\eea
by utilizing the properties of isosceles triangle. It is easy to verify that $V_{12}({\bf r}_{21},\theta)+V_{R3}(\rho,\theta)=V({\bf r}_{31})+V({\bf r}_{32})$ as should be. 

Finally, by following \eqref{Rr}, \eqref{SER} and \eqref{SEr}, the Schr$\ddot{\text o}$dinger equations for the relative motion inside diquark and diquark-quark system are given by
\bea
\left[{(\hat{p}_{\rm i}^{r}+{Q\over8}F_{\rm i\nu}{r}_{21}^\nu)^2\over 2\mu}+V({\bf r}_{21})+V_{12}({\bf r}_{21},\theta)-E_{\rm r}^{12}\right]\psi({\bf r}_{21})&=&0,\label{SE21}\\
\left[{(\hat{p}_{\rm i}^{\cal R}-{\cal Q}A_{\rm i}(\mathbb{{\cal R}}))^2\over 2{\cal M}}+{(\hat{p}_{\rm i}^{\rho}+{q\over3}F_{\rm i\nu}{\rho}^\nu)^2\over 2{\mu'}}+V_{R3}(\boldsymbol{\rho},\theta)-(E-E_{\rm r}^{12})\right]{\Psi}(\mathbb{{\cal R}},\boldsymbol{\rho})&=&0.
\eea
\end{widetext}
In the second equation, we find the reduced charge for the relative motion between quark and diquark to be ${2\over 3}q$, and the relevant notations are
 \bea
 {\cal M}&\equiv& M+m_3=3m,\\ 
 {\cal Q}&=&Q+q^3=3q, \\
 \mathbb{{\cal R}}&\equiv&\tilde{M} {\bf R}+\tilde{m}_3{\bf r}_3={2\over3} {\bf R}+{1\over3}{\bf r}_3,\\
  {\mu'}&=&\left({ M}^{-1}+{m_3}^{-1}\right)^{-1}={2\over3}m.
 \eea
Again, the reduced charge and mass are found to be evaluated in the same way for the quark-diquark system. If we redefine ${\Psi}(\mathbb{{\cal R}},\boldsymbol{\rho})={\varphi}(\boldsymbol{\rho}){\Phi}(\mathbb{{\cal R}})$, then two independent equations follow for the diquark-quark system as
\bea
\left[{(\hat{p}_{\rm i}^{\cal R}-{\cal Q}A_{\rm i}(\mathbb{{\cal R}}))^2\over 2{\cal M}}-(E-E_{\rm r}^{12}-E_\rho)\right]{\Phi}(\mathbb{{\cal R}})&=&0,\label{SE}\\
\left[{(\hat{p}_{\rm i}^{\rho}+{q\over3}F_{\rm i\nu}{\rho}^\nu)^2\over 2{\mu'}}+V_{R3}(\rho,\theta)-E_\rho\right]{\varphi}(\boldsymbol{\rho})&=&0.\label{SE3R}
\eea
In principle, the second equation is two dimensional as the constraint $\boldsymbol{\rho}\cdot {\bf r}_{21}=0$ should be satisfied. But we will firstly solve \eqref{SE21} and \eqref{SE3R} independently for a given $\theta$ and then apply the constraint by requiring a self-consistent condition, $2\tan\theta=\langle r_{21}\rangle/\langle \rho \rangle$. We call this "quark-diquark isosceles (QDI)" approximation. Actually, \eqref{SE21} and \eqref{SE3R} are the same when we set $\theta={\pi\over6}$. For the eigenstates with the same quantum numbers, the relative energy is $2E_{\rm r}^{12}$, and the equilateral triangle structure remains the self-consistent solution despite the anisotropy introduced by the magnetic field. 

There is another simpler approximation. If the diquark is tightly bound, that is, $|{\bf r}_{21}|\ll |\boldsymbol{\rho}|$, we can set $V({\bf r}_{31})=V({\bf r}_{32})=V\left(\boldsymbol{\rho}\right)$. Then, the independent Schr$\ddot{\text o}$dinger equations would follow in a similar way as the QDI approximation as
\bea
\left[{(\hat{p}_{\rm i}^{\cal R}-{\cal Q}A_{\rm i}({{\cal R}}))^2\over 2{\cal M}}-(E-E_{\rm r}^{21}-E_\rho)\right]{\Phi}({{\cal R}})&=&0,\label{TD0}\\
\left[{(\hat{p}_{\rm i}^{r_{21}}\!-\!{q\over2}A_{\rm i}({\bf r}_{21}))^2\over m}\!+\!V\left({\bf r}_{21}\right)\!-\!E_{\rm r}^{21}\right]\!{\psi}({\bf r}_{21})&=&0,\label{TD1}\\
\left[{(\hat{p}_{\rm i}^{\rho}+{q\over3}F_{\rm i\nu}{\rho}^\nu)^2\over 2{\mu'}}+2V(\rho)-E_\rho\right]{\varphi}(\boldsymbol{\rho})&=&0,\label{TD2}
\eea
which are all three dimensional. One should be cautious that these equations only give correct solutions to baryons when the constraint $\langle r_{21}\rangle\ll\langle \rho \rangle$ is  self-consistently satisfied. But it is usually not the case for the lowest-lying states, hence the QDI approximation provides a more suitable and systematic way to explore these states. Note that in the small angle limit $\theta\rightarrow 0$, the relative Schr$\ddot{\text o}$dinger equations in the QDI approximation, \eqref{SE21} and \eqref{SE3R}, exactly reduce to \eqref{TD1} and \eqref{TD2}, respectively. So, the QDI approximation automatically covers the case when the diquark is tightly bound.

\subsubsection{Reductions in cylindrical coordinate system}
In the QDI approximation, if we recover the magnetic-spin coupling, the Schr$\ddot{\text o}$dinger equations,  \eqref{SE21}, \eqref{SE} and \eqref{SE3R}, become
\bea
\!\!\left[{(\hat{p}_{\rm i}^{\cal R}-{\cal Q}A_{\rm i}(\mathbb{{\cal R}}))^2\over 2{\cal M}}-(E-E_{\rm r}-E_\rho)\right]{\Phi}&=&0,\label{SCM}
\\
\!\!\left[{3(\hat{p}_{\rm i}^{\rho}+{q\over3}F_{\rm i\nu}{\rho}^\nu)^2\over 4m}+V(\rho,\theta)-{qS_{\rm z}B\over m}-E_\rho\right]{\varphi}&=&0,\label{Sqd}
\\
\!\!\left[{(\hat{p}_{\rm i}^{r}+{Q\over8}F_{\rm i\nu}{r}^\nu)^2\over m}+V({\bf r})+V({\bf r},\theta)-E_{\rm r}\right]\psi&=&0.\label{Sd}
\eea
Here, the scripts ${21}$ and ${R3}$ are suppressed for simplicity without causing confusion, and $S_{\rm z}\equiv \sum_{i=1,2,3}S_{\rm iz}$ is the spin of baryon along the magnetic field with $S_{\rm iz}$ the quark spin. Since we are mainly interested in the total relative energy $E_{\rm r}+E_\rho$, we simply put all the magnetic-spin couplings of quarks together in \eqref{Sqd}. The center of mass Schr$\ddot{\text o}$dinger equations \eqref{SCM} is just the one for a free particle in a constant magnetic field, thus the eigenenergy can just be given with the help of Landau levels as
\bea
(E-E_{\rm r}-E_\rho)={(2n+1)|{\cal Q}B|+k_{\rm z}^2\over 2{\cal M}},\ n\in\mathbb{N}.\label{ECM}
\eea
Compared to mesons, $S_{\rm z}B$ would not change the spin structure of a single-flavor baryon, thus the magnetic-spin coupling could altogether be put into the quark-diquark relative Schr$\ddot{\text o}$dinger equation \eqref{Sqd}. Since $S={1/ 2}$ or ${3/ 2}$, we always expect the Zeeman splitting effect for baryons with different $S_{\rm z}$ components.

Now, we try to decrease the number of variables in the relative Schr$\ddot{\text o}$dinger equations by referring to a cylindrical coordinate system with 
azimuthal angles  and transverse radii defined as
\bea
\phi_\rho&=&\arctan{\rho_{\rm y}\over \rho_{\rm x}},\ \eta_\rho=\sqrt{\rho_{\rm x}^2+\rho_{\rm y}^2};\\
\phi_{\rm r}&=&\arctan{r_{\rm y}\over r_{\rm x}},\ \ \eta_{\rm r}=\sqrt{r_{\rm x}^2+r_{\rm y}^2}.
\eea
With that, the relative Schr$\ddot{\text o}$dinger equations \eqref{Sqd} and \eqref{Sd} become
\bea
 \bigg[-\frac{3}{4m}\left(\frac{\partial^2}{\partial\eta_\rho^2}+\frac{1}{\eta_\rho}\frac{\partial}{\partial\eta_\rho}+\frac{1}{\eta_\rho^2}\frac{\partial^2}{\partial\phi_\rho^2}+\frac{\partial^2}{\partial \rho_{\rm z}^2}\right)&-&\frac{iqB}{2m}\frac{\partial}{\partial\phi_\rho}\nonumber \\
 + \frac{q^2B^2\eta_\rho^2}{12m}+V(\rho,\theta)-{qS_{\rm z}B\over m}-E_\rho\bigg]{\varphi}&=&0,\\
\bigg[-\frac{1}{m}\left(\frac{\partial^2}{\partial\eta_{\rm r}^2}+\frac{1}{\eta_{\rm r}}\frac{\partial}{\partial\eta_{\rm r}}+\frac{1}{\eta_{\rm r}^2}\frac{\partial^2}{\partial\phi_{\rm r}^2}+\frac{\partial^2}{\partial r_{\rm z}^2}\right)&-&\frac{iqB}{2m}\frac{\partial}{\partial\phi_{\rm r}}\nonumber \\
+ \frac{q^2B^2\eta_{\rm r}^2}{16m}+V(r)+V(r,\theta)-E_{\rm r}\bigg]\psi&=&0.
\eea
The potentials do not explicitly depend on the azimuthal angles, so we could present the wave functions in the eigenstates of orbital angular momentum, 
\bea
{\varphi}(\boldsymbol{\rho})&\equiv& {e^{i\,l_\rho \phi_\rho}\over \sqrt{\eta_\rho}}{\varphi}_{l_\rho}(\eta_\rho, z_\rho),\\
{\psi}({\bf r})&\equiv& {e^{i\,l_{\rm r} \phi_{\rm r}}\over \sqrt{\eta_{\rm r}}}{\psi}_{l_{\rm r}}(\eta_{\rm r}, z_{\rm r}),
\eea
and find reduced forms as
\bea
\!\!\!\bigg[-\frac{3}{4m}\left(\frac{\partial^2}{\partial\eta_\rho^2}+\frac{1/4-l_\rho^2}{\eta_\rho^2}+\frac{\partial^2}{\partial \rho_{\rm z}^2}\right)+\frac{qBl_\rho}{2m}&&\nonumber \\
+ \frac{q^2B^2\eta_\rho^2}{12m}+V(\rho,\theta)-{qS_{\rm z}B\over m}-E_\rho\bigg]&&{\varphi}_{l_\rho}=0,\label{SErho}\\
\!\!\! \bigg[-\frac{1}{m}\left(\frac{\partial^2}{\partial\eta_{\rm r}^2}+\frac{1/4-l_{\rm r}^2}{\eta_{\rm r}^2}+\frac{\partial^2}{\partial r_{\rm z}^2}\right)+\frac{qBl_{\rm r}}{2m}&&\nonumber \\
 + \frac{q^2B^2\eta_{\rm r}^2}{16m}+V(r)+V(r,\theta)-E_{\rm r}\bigg]&&{\psi}_{l_{\rm r}}=0.\label{SEr}
\eea
These Schr$\ddot{\text o}$dinger equations are effectively two dimensional and could be solved numerically and independently for given physical parameters, $T, \mu_{\rm B}$ and $eB$, vertex angle $\theta$, and angular quantum numbers, $l_\rho$ and $l_{\rm r}$. Then, the total relative energy of the baryon is given by $E_{\rm t}\equiv E_{\rm r}+E_\rho$. Note that $E_{\rm t}$ cannot be taken as the binding energy of the baryon here by following the QED case, because confining effect is involved in a QCD system and no free quarks can be separated out. Instead, it is better to define the total relative energies of single-flavor heavy baryons by $E_{\Omega_{ccc}}=3m_D-m_{\Omega_{ccc}}$ and $E_{\Omega_{bbb}}=3m_B-m_{\Omega_{bbb}}$.

Similarly, when the diquark is tightly bound, the relative Schr$\ddot{\text o}$dinger equations, \eqref{TD1} and \eqref{TD2}, can be reduced to
\bea
\!\!\!\bigg[-\frac{3}{4m}\left(\frac{\partial^2}{\partial\eta_\rho^2}+\frac{1/4-l_\rho^2}{\eta_\rho^2}+\frac{\partial^2}{\partial \rho_{\rm z}^2}\right)+\frac{qBl_\rho}{2m}&&\nonumber \\
+ \frac{q^2B^2\eta_\rho^2}{12m}+2V(\rho)-{qS_{\rm z}B\over m}-E_\rho\bigg]&&{\varphi}_{l_\rho}=0,\\
\bigg[-\frac{1}{m}\left(\frac{\partial^2}{\partial\eta_{\rm r}^2}+\frac{1/4-l_{\rm r}^2}{\eta_{\rm r}^2}+\frac{\partial^2}{\partial r_{\rm z}^2}\right)+\frac{qBl_{\rm r}}{2m}&&\nonumber \\
+ \frac{q^2B^2\eta_{\rm r}^2}{16m}+V(r)-E_{\rm r}\bigg]&&{\psi}_{l_{\rm r}}=0.
\eea

\section{Numerical results}\label{num}
Here, we present numerical results for $\Omega_{\rm ccc}$ and $\Omega_{\rm bbb}$ with the component quark masses and charges, respectively,
\bea
m_{\rm c}&=&1.29~{\rm GeV},\ q_{\rm c}={2\over3}e;\\
m_{\rm b}&=&4.70~{\rm GeV},\ q_{\rm b}=-{1\over3}e.
\eea
Their total spin is $S= {3\over2}$, thus the $z$-component could be $S_{\rm z}=\pm {1\over2}, \pm {3\over2}$. And the orbital angular momenta $l_{\rm r}$ and $l_\rho$ should be even in order to keep the total wave functiolossn anti-commutative when exchanging any two quarks. In the following, we set $l_{\rm r}=l_\rho=0$ to study the lowest-lying eigenstates and the QDI approximation will be taken.

\begin{figure}[!htb]
	\begin{center}
	\includegraphics[width=8cm]{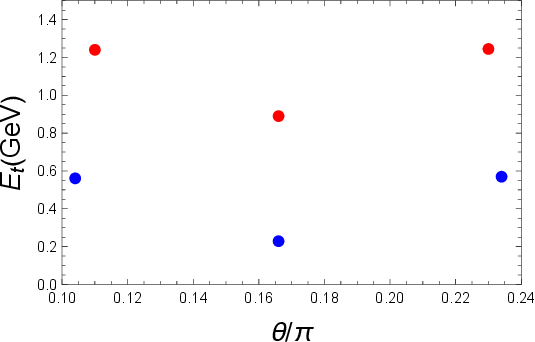}
	\caption{A few lowest-lying eigenenergies $E_{\rm t}$ of $\Omega_{\rm ccc}$ (red)  and $\Omega_{\rm bbb}$ (blue) as functions of the corresponding self-consistent vertex angles $\theta$ in the vacuum. }\label{Ebth}
	\end{center}
\end{figure}

In the vacuum with $m_{\rm D}=0$ and $eB=0$, three lowest-lying eigenenergies $E_{\rm t}$ of $\Omega_{\rm ccc}$ and $\Omega_{\rm bbb}$ are demonstrate as functions of the corresponding self-consistent vertex angles $\theta$ in Fig.~\ref{Ebth}. As we can see, the eigenstates take widely separated values of $\theta$ due to the quantization, and the ground and first excitation states are singlet and doublet in coordinate space, respectively. Note that there could be degenerate states from the cases with $l_{\rm r}\neq 0$ or $l_\rho\neq 0$ for the excitation states, while the ground states are usually singlet given by $l_{\rm r}=l_\rho= 0$~\cite{Xiang:2024wxh}. As a check of the validity of the method, the binding energies of the ground states are found to be $E_{\Omega_{ccc}}=0.964~{\rm GeV}$ and $E_{\Omega_{\rm bbb}}=1.60~{\rm GeV}$, in good agreement with those given in the literatures~\cite{Brown:2014ena,He:2014tga,Zhao:2023qww}.

\begin{figure}[!htb]
	\begin{center}
	\includegraphics[width=8cm]{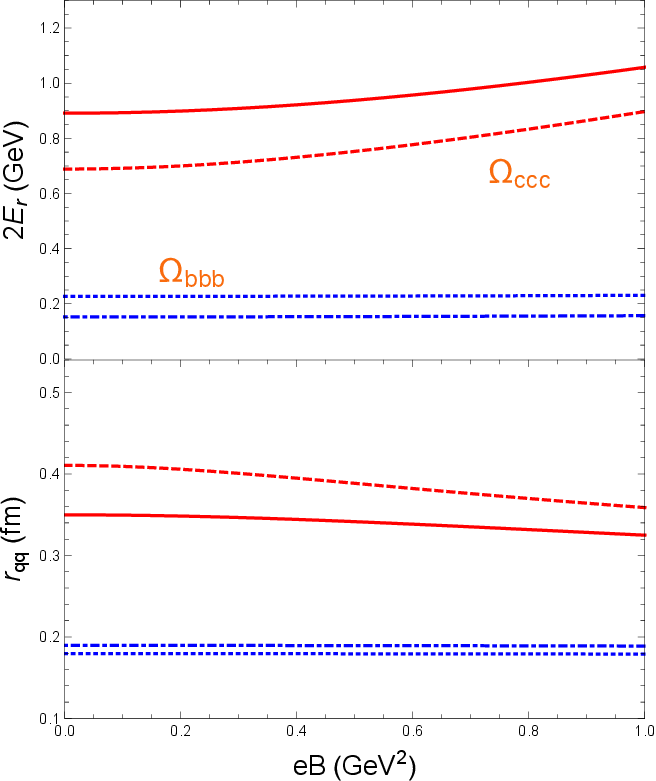}
	\caption{The lowest-lying orbital eigenenergies $2E_{\rm r}$ and two-quark average distances $r_{\rm qq}$ of $\Omega_{\rm ccc}$ (red)  and $\Omega_{\rm bbb}$ (blue) as functions of the magnetic field $eB$ for $m_{\rm D}=0$ (solid and dotted) and $0.4~{\rm GeV}$ (dashed and dashdotted). }\label{pB}
	\end{center}
\end{figure}

\begin{figure}[!htb]
	\begin{center}
	\includegraphics[width=8cm]{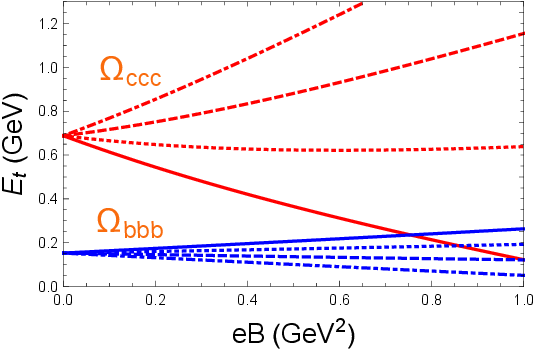}
	\caption{The total relative energies $E_{\rm t}$ of $\Omega_{\rm ccc}$ (red) and $\Omega_{\rm bbb}$ (blue) as functions of the magnetic field $eB$ in the quark-gluon-plasma phase with spins $S_{\rm z}= {3\over2}$ (solid), $ {1\over2}$ (dotted), $-{1\over2}$ (dashed), and $-{3\over2}$ (dashdotted). The orbital eigenenergies correspond to those given in Fig.~\ref{pB} for $m_{\rm D}=0.4~{\rm GeV}$.}\label{pBs}
	\end{center}
\end{figure}

Next, we explore the magnetic effect to the ground state when QCD medium is absent ($m_{\rm D}= 0$) or present ($m_{\rm D}\neq 0$). As mentioned before, we cannot tell the difference between the effects of temperature and baryon density once $m_{\rm D}$ is given. In the following, we will ignore the concrete values of temperature and baryon density, but simply take $m_{\rm D}=0.4~{\rm GeV}$ to stand for the quark-gluon plasma phase. For the ground state, the quantum numbers are the same for the lowest-lying eigenstates to \eqref{SErho} and \eqref{SEr}, hence $\theta={\pi\over 6}$ and the relative energy is given by $E_{\rm t}=2E_{\rm r}-qS_{\rm z}B/m$. The results without and with magnetic-spin coupling are given in Fig.~\ref{pB} and Fig.~\ref{pBs}, respectively. 

From Fig.~\ref{pB}, it is easy to find that $\Omega_{\rm ccc}$ is more sensitive to the magnetic field, since both the two-quark average distance $r_{\rm qq}$ and absolute charge are larger compared to $\Omega_{\rm bbb}$. For $\Omega_{\rm ccc}$, $r_{\rm qq}$ decreases with $eB$ while the lowest-lying orbital eigenenergy $2E_{\rm r}$ increases, that is, the baryon is more tightly bound in the presence of magnetic field. These features can be well understood by noticing that the magnetic field only functions through the quadratic terms $\sim{q^2B^2}$ in \eqref{SErho} and \eqref{SEr} for $l_{\rm r}=l_\rho=0$. For $\Omega_{\rm bbb}$, the orbital properties are almost not affected by $eB$ in the considered region. The reason is that the effective magnetic length, evaluated as $l_{\rm B}\equiv \sqrt{2/|q_{\rm b}B|}>0.48~{\rm fm}$ according to \eqref{SE21}, is larger than the two-quark average distance $r_{\rm qq}$ for $\Omega_{\rm bbb}$. 

When the magnetic-spin couplings are taken into account in the quark-gluon-plasma phase, Zeeman splitting effect immediately shows up for both baryons. According to Fig.~\ref{pBs}, the orbital eigenenergies $2E_{\rm r}$ changes with the magnetic field approximately as $ {|qB|\over 2m}$, and the magnetic-spin effect dominates over the magnetic-orbital effect for the lowest-lying states, that is, with $S_{\rm z}= {3\over2}$ for $\Omega_{\rm ccc}$ and $S_{\rm z}= -{3\over2}$ for $\Omega_{\rm bbb}$.

\section{Summary}\label{summary}
In this work, a systematic formalism is developed to deal with the two-body Schr$\ddot{\text o}$dinger equation in the presence of a constant magnetic field. Such a formalism is very useful for a single-flavor system since the center of mass and relative motions can be completely separated from each other. Then, we apply the formalism to deal with a single-flavor heavy baryon, which is simply considered as a quark-diquark two-body system with diquark itself a quark-quark two-body system. Assuming the quarks form an isosceles triangle on average, a self-consistent method is proposed to study such a baryon. The method is quite optimistic since it is capable to reproduce both the isotropic state in the vanishing magnetic field limit and cover the case when the diquark is tightly bound in the small angle limit.

Eventually, the method is applied to two triply heavy baryons, $\Omega_{\rm ccc}$ and $\Omega_{\rm bbb}$ -- the validity is justified by the good agreement between our predictions and other studies on the lowest-lying eigenenergies in the vacuum. The magnetic effect to the ground state is explored when QCD medium is absent or present, and it is found that: 1. The orbital properties of $\Omega_{\rm bbb}$ are almost not affected by $eB$ since both the absolute charges and two-quark average distance are small. 2. $\Omega_{\rm ccc}$ is more tightly bound in the presence of magnetic field as the two-quark average distance decreases with $eB$. 3. The magnetic-spin effect dominates over the magnetic-orbital effect for both the lowest-lying states of $\Omega_{\rm ccc}$ and $\Omega_{\rm bbb}$. As a conclusion, to study the magnetic effect, $\Omega_{\rm ccc}$ is a much better probe than $\Omega_{\rm bbb}$. Moreover, recalling the center of mass eigenenergy in \eqref{ECM}, the lowest energy of $\Omega_{\rm ccc}$ changes with the magnetic field as 
\bea
\Delta E_0={|QB|\over 2{\cal M}}-{3|qB|\over 2m}+2\Delta E_{\rm r}\approx -{|qB|\over 2m}.
\eea
Thus, the lowest energy $E_0$ decreases with $|eB|$ and the discovery of $\Omega_{\rm ccc}$ could be more promising in peripheral heavy ion collisions.

\section*{Acknowledgment}
G. C. thanks Jiaxing Zhao and Shuzhe Shi for helpful communications. G. C. is funded by both the "Forefront Leading Project of Theoretical Physics" from the National Natural Science Foundation of China with Grant No. 12447102. and the Natural Science Foundation of Guangdong Province with Grant No. 2024A1515011225.

\end{document}